\documentclass[useAMS,usenatbib]{mn2e}
\usepackage{graphicx}

\defcitealias{FFG07}{FFG07}
\defcitealias{THO04}{TMK04}

\title[Multicolour-Metallicity Relations from Globular Clusters in
NGC\,4486]{Multicolour-metallicity Relations from Globular Clusters in
  NGC\,4486 (M87)%
\thanks{Based on observations obtained at the Gemini Observatory, which is
    operated by the Association of Universities for Research in Astronomy,
    Inc., under a cooperative agreement with the NSF on behalf of the Gemini
    partnership: the National Science Foundation (United States), the
    Science and Technology Facilities Council (United Kingdom), the National
    Research Council (Canada), CONICYT (Chile), the Australian Research
    Council (Australia), Ministerio da Ciencia e Tecnologia (Brazil) and
    Ministerio de Ciencia, Tecnolog\'ia, e Innovaci\'on Productiva
    (Argentina).}}

\author[Forte et al.]{Juan
  C. Forte$^{1,2}$\thanks{E-mail: forte@fcaglp.unlp.edu.ar}, Favio
  R. Faifer$^{2,3,4}$, E. Irene Vega$^{2,3}$, Lilia P. Bassino$^{2,3,4}$,
  \newauthor 
 Anal\'ia V. Smith Castelli$^{2,3,4},$ Sergio A. Cellone$^{2,3,4}$, Douglas
 Geisler$^{5}$\\
$^1${Planetario ``Galileo Galilei'', Secretar\'ia de Cultura, Ciudad
  Aut\'onoma de Buenos Aires, Argentina}\\
$^2${Consejo Nacional de Investigaciones Cient\'ificas y T\'ecnicas,
  Av. Rivadavia 1917, C1033AAJ, Ciudad Aut\'onoma de Buenos Aires,
  Argentina}\\
$^3${Facultad de Ciencias Astron\'omicas y Geof\'isicas, Universidad
  Nacional de La Plata, Paseo del Bosque, B1900FWA, La Plata,
  Argentina}\\
$^4${Instituto de Astrof\'isica de La Plata (CCT-La Plata, CONICET-UNLP),
  Paseo del Bosque, B1900FWA, La Plata, Argentina}\\
$^5${Departamento de Astronom\'ia, Universidad de Concepci\'on, Casilla
  160-C, Concepci\'on, Chile}
}

\begin{document}

\date{Accepted ;  Received ; in original form }
\pagerange{\pageref{firstpage}--\pageref{lastpage}} \pubyear{}
\maketitle

\label{firstpage}

\begin{abstract}
                                                                       
  We present Gemini $griz'$ photometry for 521 globular cluster (GC)
  candidates in a $5.5\times 5.5$ arcmin field centred 3.8 arcmin to the
  south and 0.9 arcmin to the west of the centre of the giant elliptical
  galaxy NGC\,4486. All these objects have previously published $(C-T_1)$
  photometry. We also present new $(C-T_1)$ photometry for 338 globulars, 
  within 1.7 arcmin in galactocentric radius, which have $(g-z)$ colours in 
  the photometric system adopted by the Virgo Cluster Survey of the Advanced
  Camera for Surveys of the Hubble Space Telescope (HST).
  These photometric data are used to define a self-consistent multicolour
  grid (avoiding polynomial fits) and preliminarily calibrated in terms of 
  two chemical abundance scales. The resulting multicolour colour-chemical 
  abundance relations are used to test GC chemical abundance distributions. 
  This is accomplished by modelling the ten GC colour histograms that can be 
  defined in terms of the ${\it Cgriz'}$ bands. Our results suggest that the 
  best fit to the GC observed colour histograms is consistent with a genuinely 
  bimodal chemical abundance distribution $N_\mathrm{GC}(Z)$. On the other 
  side, each (``blue'' and ``red'') GC subpopulation follows a distinct 
  colour-colour relation.

\end{abstract}
\section{Introduction}
\label{INTRO}

Globular clusters (GCs) are tracers of early events in the star forming
history in galaxies. However, a unique integrating picture of that history,
beyond some tentative approaches, is still missing.  A thorough review of
several issues in this context is presented, for example, in
\citet{BRO06}. Important aspects, that eventually deal with large scale
properties of galaxies \citep*[see, for example][and references
therein]{FVF12} are both the age and chemical abundance distribution of
these clusters.

  Even though the quality and volume of chemical abundance ([Z/H])
  data for GCs is steadily growing (\citealt{ALV11}; \citealt{USH12}), a
  key issue remains as an open subject: the connection between the GC
  abundances and their integrated colours. 

  Under the common assumption of old ages, GC integrated colours should
  be dominated by chemical abundance (and in a secondary way by age).
  Evidence in this sense can be found, for example, in \citet{NOR08} and,
  in the particular case of elliptical galaxies, in \citet{CHI12}.

  A survey of the literature reveals numerous attempts to link colours
  and chemical abundance, ranging from linear \citep{GEI90}, 
  quadratic (\citealt{HAR02}; \citealt*{FFG07}; \citealt*{MOY10}) or quartic
  dependences \citep*{BLA10}. A recent contribution on this subject
  has been presented by \citet{USH12} who adopt broken line fits.

 A clarification of the colour-abundance connection is required, since
 some non linear relations (e.g. ``inflected") can eventually lead to
 GC bimodal colour distributions, even when a unimodal chemical abundance 
 distribution is assumed \citep[see, for example,][]{YOO11}.
 Since bimodal colour distributions have {\bf often} been identified
 as the result of genuine bimodal chemical abundance distributions, the
 presence of non linearities would have important
 consequences on the interpretation of the GC chemical abundance
 distributions and also on their possible quantitative connections with
 the diffuse stellar population in a given galaxy.


 NGC\,4486 is a particularly useful galaxy in order to revise the chemical
 abundance-integrated colour issue due to its large GC population and
 relative proximity to the Sun, $\approx 16.6$ Mpc (\citealt*{TON01}; \citealt*{BLA09}).
 This paper presents Gemini $griz'$ high quality photometry for a selected field
 centred 3.9 arcmin from the centre of the galaxy, including 521 GC
 candidates, and new $(C-T_1)$ photometry for 338 clusters within a
 galactocentric radius of 1.7 arcmin, which also have $(g-z)$ colours
 obtained with the Advanced Camera for Surveys (ACS) of the Hubble Space
 Telescope (HST) \citep{JOR09}. In addition, aiming at extending
 the wavelength coverage towards the ultraviolet, we combine our Gemini
 photometry with the $C$ magnitudes \citep[Washington system;][]{HARC77}
 data presented by \citet[hereafter FFG07]{FFG07}. All these data sets can
 be, firstly, mutually connected to define a self consistent multicolour
 grid, and then calibrated in terms of different chemical abundance scales,
with a final goal of obtaining a simultaneous connection between metallicity and the
 colour indices grid.

 The structure of the paper is as follows. Observations and data handling
 are presented in Section\,\ref{OBS}. The relation between the ten colour
 indices defined by the ${\it Cgriz'}$ photometry is given in
 Section\,\ref{MULTICOLOUR}.  This last section also explains the connection
 between those colour indices and others, like $(C-T_1)$,
 $(g-z)_\mathrm{(ACS)}$, commonly used in extragalactic
 GCs research.  Section\,\ref{COLCHEM} presents a preliminary calibration
 in terms of chemical abundance using two different empirical colour-metallicity
 scales: \citet{BLA10}, and alternatively, the scale presented by 
 \citet{USH12}, that are refreered to as B10 and U12, respectively, in what 
 follows. As explained in Section\,\ref{GCs}, the multicolour grid can be 
 used to define different magnitude curves (or pseudo-spectral distributions) 
 as a function of $[Z/H]$. These \emph{template} curves are used to clean the 
 GC candidates photometric sample from field interlopers. The analysis of the 
 residuals, in turn, would allow the detection of eventual age effects. The 
 determination of GCs $[Z/H]$, via  multicolor fits,  is presented in Section 
 \ref{INVERSE}. The results of confronting the ten observed GC colour 
 histograms with models that adopt each chemical abundance calibration
 are described in Section\,\ref{MODELL}. The final conclusions are given in 
 Section\,\ref{CONCLU}.


\section{Observations and data handling}
\label{OBS}
\begin{figure}
\includegraphics[width=\hsize]{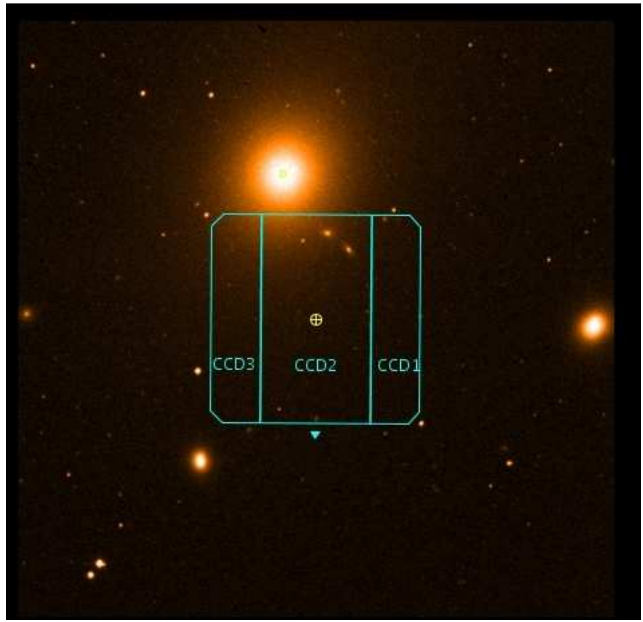}
\caption{The GMOS field (cyan lines square; 5.5 arcmin on a side) discussed 
 in this work. NGC\,4486 appears up and to the left. North is up and East to
 the left.}
\label{GMOS}
\end{figure}

The photometric observations presented in this paper were carried out with
the Fred Gillette 8-m Telescope (Gemini North) and are part of a program
that also includes GMOS spectroscopy of a sample of selected objects that
are considered as good GCs candidates (GN-2010A-Q-21; PI: J.\,C. Forte).
The field (5.5 arcmin on a side), centred 3.8 arcmin to the South and 0.9
arcmin to the West of NGC\,4486, is shown in Figure\,\ref{GMOS}.  The
observing log, including dates, filters, exposures, mean air mass and
composite seeing (FWHM) is given in Table\,\ref{Table_1}.


\begin{table}
\center
\caption{Observing log of Gemini-GMOS observations.}
\label{Table_1}
\begin{tabular}{@{}lcccc@{}}                                 
\hline
                   Date    & Filter & Exposure  &  Airmass  &  Seeing \\
                           &        &  (sec.)   &           &  (arcsec)\\
\hline
   2010/01/16     &  $g'$   &    $5 \times 400$   &   1.009   &    0.63 \\
   2010/01/16-17  &  $r'$   &    $5 \times 300$   &   1.017   &    0.60 \\
   2010/01/17     &  $i'$   &    $5 \times 300$   &   1.147   &    0.54 \\
   2010/01/17     &  $z'$   &    $6 \times 450$   &   1.080   &    0.47 \\
\hline
\end{tabular}
\end{table}

Image processing was performed with the tasks of the package \textsc{gmos}
within \textsc{iraf}\footnote{\textsc{iraf} is distributed by the National
  Optical Astronomy Observatories, which are operated by the Association of
  Universities for Research in Astronomy, Inc., under cooperative agreement
  with the National Science Foundation.}. In turn, PSF magnitudes were
obtained with the package \textsc{daophot} within \textsc{iraf}.  These
instrumental magnitudes were corrected for atmospheric extinction adopting
the coefficients given in the Gemini web
pages\footnote{{http://www.gemini.edu/sciops/instruments/gmos/calibration?}
{q=node/10445}},
and transformed to the $griz'$ system using zero points derived from
observations of a standard field (PG1323$+$086), in a way similar to that
extensively described in \citet{FAI11}.  The identification of GC
candidates, using both \textsc{daophot} and \textsc{SExtractor}
\citep{BER96}, also follows the lines described in that paper.
The limiting magnitude of the photometric sample is $g'_0 \approx 23.7$ mag,
i.e., $\approx 0.15$ mag brighter than the turn-over of the GCs integrated
luminosity function \citep{VILL10}. The photometric errors as a function of
the $g'$ magnitudes are given in Table\,\ref{Table_2}.
     
In what follows we adopt a colour excess $E(B-V)=0.02$ mag from the maps by
\citet*{SCH98}, and the interstellar extinction relations given in
\citet{JOR04}. The magnitudes and colours, corrected for interstellar
reddening (denoted with the ``$0$'' subscript), of our GC sample are given in
Table\,\ref{Table_3}.
 

\begin{table}
\center
\label{Table_2}
\caption{$Cgriz'$ photometry errors, in magnitudes.}
\begin{tabular}{@{}cccccc@{}}                                 
\hline
$g'$ & $\sigma_{g'}$ & $\sigma_{r'}$ & $\sigma_{i'}$ & $\sigma_{z'}$ & $\sigma_C$\\
\hline
 20.5   &  0.011    &    0.010    &     0.012     &     0.019   &     0.034 \\
 21.5   &  0.012    &    0.010    &     0.012     &     0.020   &     0.033 \\
 22.5   &  0.015    &    0.011    &     0.014     &     0.020   &     0.045 \\
 23.5   &  0.027    &    0.017    &     0.022     &     0.035   &     0.055 \\
 24.0   &  0.032    &    0.021    &     0.027     &     0.039   &     0.070 \\
\hline
\end{tabular}
\end{table}

The 521 GC candidates listed in Table\,\ref{Table_3} also have $(C-T_1)$
photometry obtained with the CTIO Blanco 4-m Telescope and presented in
\citetalias{FFG07}. The GCs identification numbers are from that work.  Due
to severe incompleteness effects, data for GCs closer to the galaxy centre were
not published in \citetalias{FFG07}. Among these objects we identified
338 that were also observed in the ACS $(g-z)$ colour \citep{JOR09}. The
photometric values for these GC candidates are listed in
Table\,\ref{Table_4} where the R.A. and Dec values come from this last work.


\begin{table*}
\center
\begin{minipage}{185mm}
  \caption{Multicolour photometry for GC candidates in the NGC\,4486
    field. Magnitudes and colours are corrected for interstellar extinction
    and reddening. Identification numbers are from \citet{FFG07}. \emph{The
      full version of the table will be available in the electronic edition
      of the Journal.}}
\label{Table_3}
\setlength\tabcolsep{1.35mm}
\begin{tabular}{@{}cccccccccccccc@{}}
\hline
       ID & $g'_0$      & $(C-g')_0$ & $(C-r')_0$
          & $(C-i')_0$  & $(C-z')_0$ & $ (g-r)'_0$ & $(g-i)'_0$ & $(g-z)'_0$ 
          & $(r-i)'_0$  & $(r-z)'_0$ & $(i-z)'_0$  & $T_{1_0}$    & $(C-T_1)_0$ \\\\
\hline
 23314 & 22.38 &  0.62 &   1.26 &   1.50 &  1.61 &  0.64 &  0.87 &  0.99 &    0.23 &   0.35 &   0.12 &    21.62 &     1.39 \\
 23796 & 21.71 &  0.38 &   0.93 &   1.14 &  1.19 &  0.56 &  0.76 &  0.81 &    0.21 &   0.25 &   0.05 &    20.95 &     1.14 \\
\hline
\end{tabular}
\end{minipage}
\end{table*}


\begin{table*}
\center
  \caption{GC candidates with ACS and Washington Photometry within a
    galactocentric radius of 100 arcsecs in NGC\,4486. {\it The full version
      of the table will be available in the electronic edition of the
      Journal.}}
\label{Table_4}
\begin{tabular}{@{}cccccc@{}}                                 
\hline
  $\alpha$ & $\delta$ & $T_{1_0}$ & $g_0$ & $(C-T_1)_0$  & 
${(g-z)_0}_{\mathrm{(ACS)}}$   \\
(degrees) & (degrees) & (mag) & (mag) & (mag) & (mag) \\
\hline
  187.7037354 & 12.3645134 & 22.686 & 23.429 & 1.140 &   0.900 \\
  187.7002411 & 12.3649130 & 22.876 & 23.701 & 0.940 &   0.789 \\
\hline
\end{tabular}
\end{table*}

\begin{figure}
\includegraphics[width=\hsize]{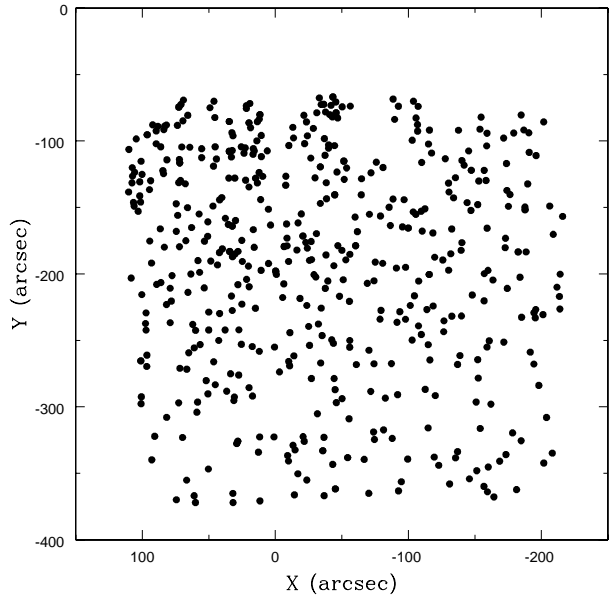}
\caption{Distribution on the sky of 521 GC candidates with ${\it Cgriz'}$
  photometry listed in Table\,\ref{Table_3}. North is up, East to the
  left. The centre of NGC\,4486 is at $x=0.0$, $y=0.0$.  }
\label{SPATIAL}
\end{figure}

\begin{figure}
\includegraphics[width=\hsize]{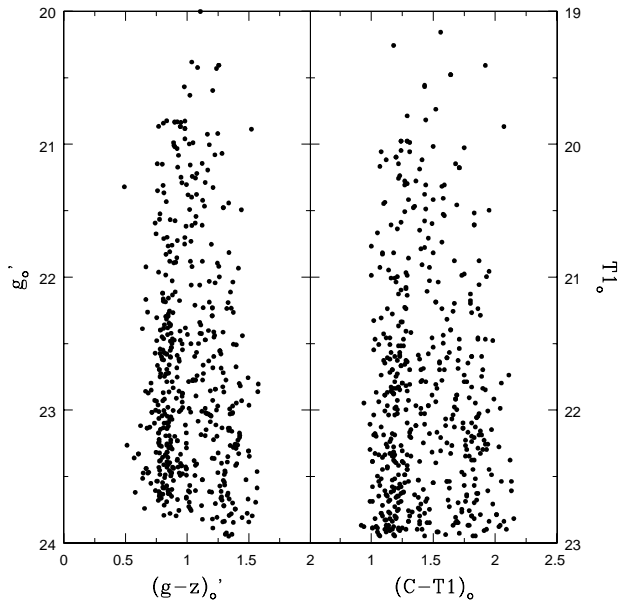}
\caption{$g'_0$ vs. $(g-z)'_0$ and $T_{1_0}$ vs. $(C-T_1)_0$ colour
  magnitude diagrams (corrected for interstellar reddening) for the GC
  candidates listed in Table\,\ref{Table_3}. Note incompleteness effects in
  the first panel for clusters bluer than ${(g-z)'}_0 =1.25$.}
\label{COLMAG}
\end{figure}

The distribution on the sky for objects within our Gemini field are depicted
in Figure\,\ref{SPATIAL}, while their $g'_0$ vs. $(g-z)'_0$ and $T_{1_0}$
vs. $(C-T_1)_0$ colour magnitude diagrams are shown in Figure\,\ref{COLMAG}.


\section{Multicolour relations}
\label{MULTICOLOUR}                                            
The adoption of a polynomial fit to represent the dependence of integrated
GC chemical abundance $[Z/H]$ with some particular colour index (or 
vice versa), is a usual approach in the literature.  For example, a linear fit
was attempted by \citet{GEI90}, while two second order polynomials,
minimizing the errors in abundance or in colour, were presented by
\citet{HAR02}.  In turn, both \citet{MOY10} and \citet{BLA10} adopted the
\textsc{robust} routine, in their respective second and quartic order
approximations.  That routine seeks minimizing the ``orthogonal'' errors
(defined in terms of a residual that combines those of the two variables).
This approach was also adopted by \citet{BLA12} when deriving the relation between
the $(g-I)$ and $(I-H)$ colours of GC candidates in the central field of
NGC\,1399.

In this paper we attempt a different approach aiming at connecting
simultaneously the ten different colour indices that can be defined through
the ${\it Cgriz'}$ photometry and avoiding polynomial fits.  The resulting
self-consistent colour-grid is then tentatively calibrated in terms of two
chemical abundance scales, as explained in the following section.

As a first step we generated nine colour-colour planes, all including the
$(g-z)'_0$ colour, the index with the longest wavelength base (in terms of
the Gemini photometry).  We preferred that index instead of $(C-z')_0$ given
the larger errors inherent to the $C$ magnitudes (see Table\,\ref{Table_2}).
Each photometric value was then convolved with a Gaussian kernel to generate
bi-dimensional images that were afterwards used to obtain modal colour
values. The adopted kernel size (0.05 mag) matches the overall error of the
colours in our sample.  

\begin{figure}
\includegraphics[width=\hsize]{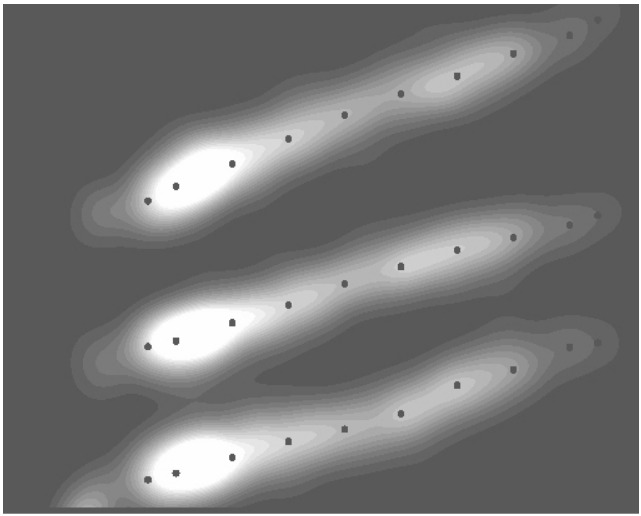}
\caption{Smoothed colour-colour relations. From top $(g-r)'_0$, $(r-i)'_0$,
  and $(i-z)'_0$ (shifted arbitrarily in ordinates) vs. $(g-z)'_0$. The
  frame is 1.1 mag on a side.}
\label{SMOOTH}
\end{figure}

These images, generated and processed with \textsc{iraf} (through the routines
 \emph{irafil} and \emph{gaussian}), were used to determine the modal values linking each index 
 at a given $(g-z)'_0$ colour (from 0.80 to 1.50 with 0.1 mag intervals, and adding
 two points at the bluest and reddest ends: 0.75 and 1.55 mag, respectively). Outside 
these limits the determination of the modes becomes uncertain.  An example,
corresponding to the $(g-r)'_0$, $(r-i)'_0$ and $(i-z)'_0$ vs. $(g-z)'_0$
relations is depicted in Figure\,\ref{SMOOTH}.

We also performed a number of simulations starting with different types of 
colour-colour relations (linear, quadratic, fourth order) and adding Gaussian 
errors (similar to those in Table\,2). The smoothing procedure was able to 
recover the input relation with a typical uncertainty ranging from $\pm$ 0.012 
mag to $\pm$ 0.02 mag. These simulations also show that changing the 
smoothing kernel from 0.03 to 0.08 mag has no detectable impact on the 
derived modal colours.


In fact, each modal colour value can be determined in four different
ways. First, directly from the plane determined by a given colour vs.
$(g-z)'_0$ and then by properly combining all the other indices including
the filter bands that define that colour (e.g., the $(C-g')_0$ vs.
$(g-z)'_0$ relation can also be determined from the $[(C-r')_0-(g-r)'_0]$,
 $[(C-i')_0-(g-i)'_0]$, and $[(C-z')_0-(g-z)'_0]$ modal colour differences
determined at a given $(g-z)'_0$). These four indices will differ as a
result of the distinct roles played by the photometric errors on each of the
two colour planes. 
 
A check for consistency indicates that, for a given mean colour, the scatter of each
four modal colours has a typical rms of $\approx 0.012$ mag 
(and $\approx 0.018$ mag in the worst case) over the whole grid. The link to 
the $(C-T_1)_0$ colour was obtained adopting the same method described above, 
i.e., by generating two smoothed colour planes and looking for the modal 
colour values. In turn, $(C-T_1)_0$ was also connected to 
${(g-z)_0}_{\mathrm{(ACS)}}$ through the GC candidates in common with 
\citet{JOR09}, listed in Table~\ref{Table_4}.

\begin{figure}
\includegraphics[width=\hsize]{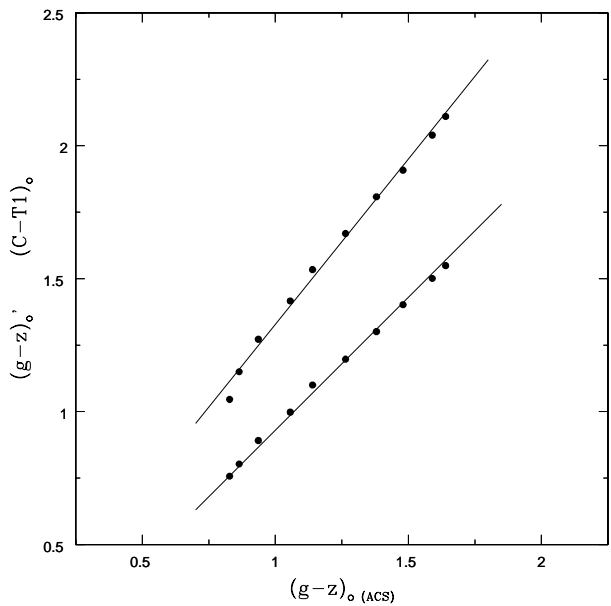}
\caption{From top to bottom: $(C-T_1)_0$ and $(g-z)'_0$
  vs.\ ${(g-z)_0}_{\mathrm{(ACS)}}$ relations. Filled dots are the modal
  colours listed in Table\,\ref{Table_5}. The straight lines correspond to
  Equation\,\ref{eqn1} and Equation\,\ref{eqn2}.}
\label{CT1GZACS}
\end{figure}

From the colour grid we derive:
\begin{equation}
\label{eqn1}
(C - T_1)_0 = 1.25\, (\pm 0.03)\, {(g-z)_0}_{\mathrm{(ACS)}} + 0.08\, (\pm 0.04).
\end{equation}
This equation corresponds to the formal bisector fit. We stress, however,
that there is a deviation at the blue extreme that, as discussed later, seems
to be a common feature in all the colour-colour relations involving the $C$
filter, possibly as a consequence of the presence of the features connected 
with the {\it 4000 \AA\, break}.
  From the same grid, we obtain:
\begin{equation}
\label{eqn2}
  (g-z)'_0 =  {(g-z)_0}_{\mathrm{(ACS)}} - 0.07 \, (\pm 0.02).
\end{equation}
The relations between $(C-T_1)_0$ and $(g-z)'_0$ with
${(g-z)_0}_{\mathrm{(ACS)}}$ given in Table\,\ref{Table_5} are displayed in
Figure\,\ref{CT1GZACS}. 


\begin{figure}
\includegraphics[width=\hsize]{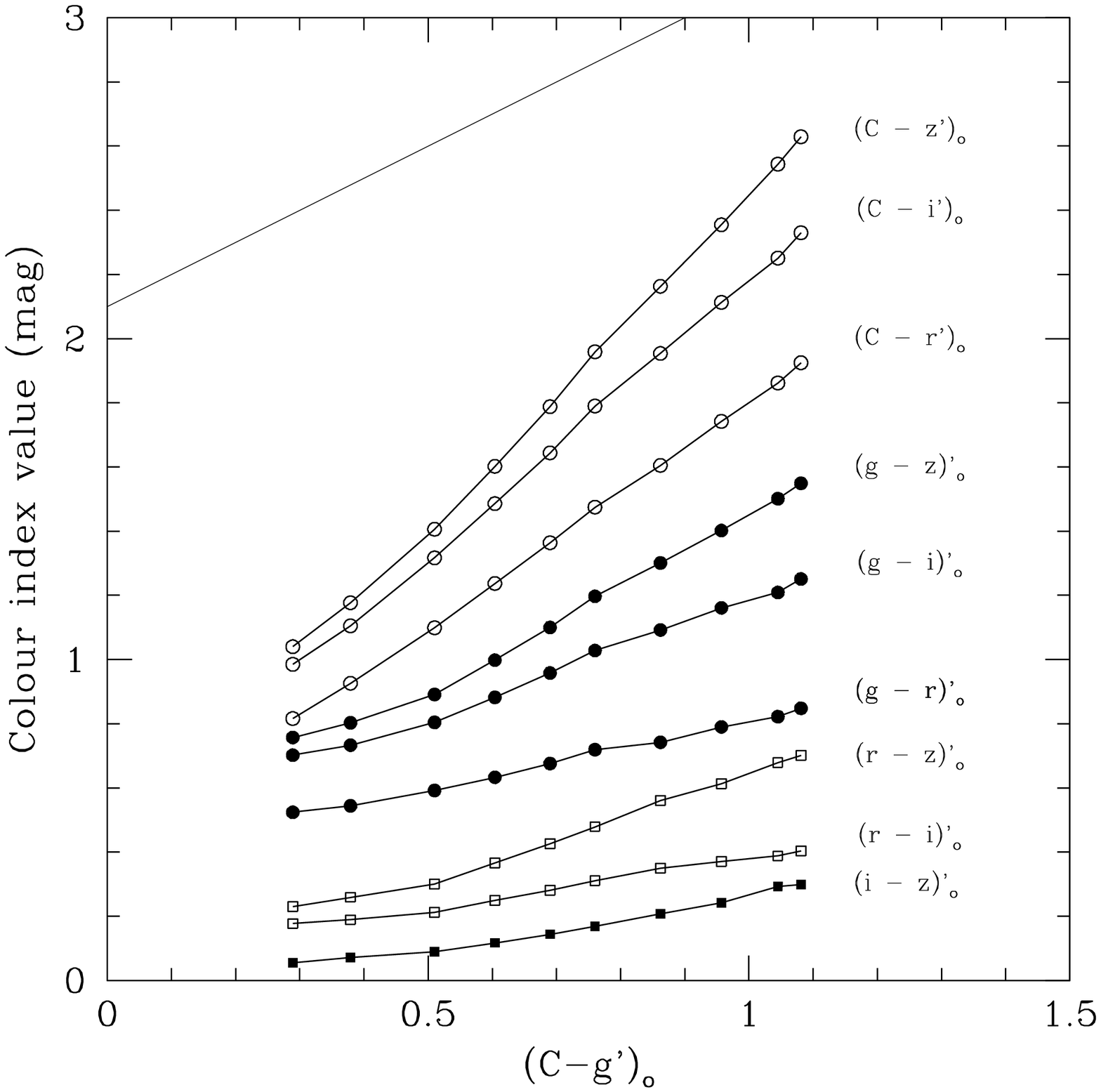}
\caption{Colour indices vs. $(C-g')_0$. From top to bottom, in decreasing
  order: $(C-z')_0$, $(C-i')_0$, $(C-r')_0$ \emph{(open dots)}; $(g-z)'_0$,
  $(g-i)'_0$, $(g-r)'_0$ \emph{(filled dots)}; $(r-z)'_0$, $(r-i)'_0$
  \emph{(open squares)}; $(i-z)'_0$ \emph{(filled squares)}. The straight
  line at upper left has a unity slope.}
\label{CGALL}
\end{figure}

\begin{figure}
\includegraphics[width=\hsize]{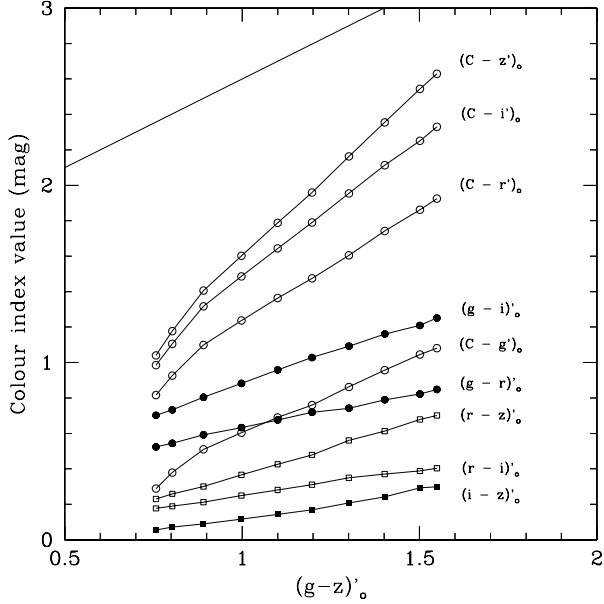}
\caption{Colour indices vs. $(g-z)'_0$. From top to bottom, in decreasing
  order: $(C-z')_0$, $(C-i)'_0$, $(C-r')_0$, $(C-g')_0$ \emph{(open dots)};
  $(g-i)'_0$, $(g-r)'_0$ \emph{(filled dots)}; $(r-z)'_0$, $(r-i)'_0$
  \emph{(open squares)}; $(i-z)'_0$ \emph{(filled squares)}. The straight
  line at upper left has a unity slope.}
\label{GZALL}
\end{figure}

The colour-colour relations defined by Table\,\ref{Table_5} display different
degrees of non linearity. In some cases, these effects can be noticed  near the
blue and red ends of the colour-colour relations. Similar kind of non 
linearities were already noticed by \citet{BLA12} in their analysis of the 
(optical) $(g-z)_\mathrm{ACS}$ vs. the (infrared) $(I-H)$ index. As examples, 
Figure\,\ref{CGALL} and Figure\,\ref{GZALL} show the behaviour of all the 
colour indices defined in our photometry as a function of $(C-g')_0$ and 
$(g-z)'_0$, respectively.


\begin{table*}
\center
\begin{minipage}{185mm}
\caption{Multicolour-Chemical abundance relations from GCs in the field of
  NGC\,4486}
\label{Table_5}
\setlength\tabcolsep{1.0mm} 
\begin{tabular}{@{}cccccccccccccc@{}}                                 
\hline
  $(C-g')_0$ & $(C-r')_0$  & $(C-i')_0$ & $(C-z')_0$ & $ (g-r)_0'$ & $(g-i)_0'$ & $(g-z)_0'$ 
              & $(r-i)_0'$  & $(r-z)_0'$ & $(i-z)_0'$ & $(C-T_1)_0$  & $(g-z)_0$  & $[Z/H]$ 
              & $[Z/H]$ \\
       & & & & & & & & & &  & ACS   & B10 & U12 \\
\hline
    0.29  &  0.82  &  0.99  &  1.04  &  0.52  &  0.70  &  0.75  &  0.18  &  0.23  &  0.05  &  1.05  &  0.83  & $-2.16$  & $-1.87$ \\
    0.38  &  0.93  &  1.11  &  1.18  &  0.54  &  0.73  &  0.80  &  0.19  &  0.26  &  0.07  &  1.15  &  0.86  & $-1.88$  & $-1.64$ \\
    0.51  &  1.10  &  1.32  &  1.41  &  0.59  &  0.80  &  0.90  &  0.21  &  0.30  &  0.09  &  1.27  &  0.94  & $-1.35$  & $-1.24$ \\
    0.60  &  1.24  &  1.49  &  1.60  &  0.63  &  0.88  &  1.00  &  0.25  &  0.37  &  0.12  &  1.42  &  1.06  & $-0.84$  & $-0.96$ \\
    0.69  &  1.36  &  1.64  &  1.79  &  0.68  &  0.96  &  1.10  &  0.28  &  0.43  &  0.14  &  1.53  &  1.14  & $-0.64$  & $-0.68$ \\
    0.76  &  1.48  &  1.79  &  1.96  &  0.72  &  1.03  &  1.20  &  0.31  &  0.48  &  0.17  &  1.67  &  1.26  & $-0.44$  & $-0.42$ \\
    0.86  &  1.61  &  1.95  &  2.16  &  0.74  &  1.09  &  1.30  &  0.35  &  0.56  &  0.21  &  1.81  &  1.38  & $-0.23$  & $-0.22$ \\
    0.96  &  1.74  &  2.11  &  2.36  &  0.79  &  1.16  &  1.40  &  0.37  &  0.61  &  0.24  &  1.91  &  1.48  & $\phantom{-}0.09$  & $\phantom{-}0.02$ \\
    1.04  &  1.86  &  2.25  &  2.54  &  0.82  &  1.21  &  1.50  &  0.39  &  0.68  &  0.29  &  2.04  &  1.59  & $\phantom{-}~0.70$:& $\phantom{-}0.19$ \\
    1.08  &  1.92  &  2.33  &  2.63  &  0.85  &  1.25  &  1.55  &  0.40  &  0.70  &  0.30  &  2.11  &  1.64  & $\phantom{-}-- $   & $\phantom{-}0.33$ \\
\hline
\end{tabular}
\end{minipage}
\end{table*}


\section{Colour-Chemical abundance relations}
\label{COLCHEM}

Despite significant effort, the shape of the
relation between GC integrated colours and chemical abundance still remains
a subject of debate.  The main problems behind a proper determination of
their relation are both the uncertainties of synthetic models and the still
large errors associated with the GC chemical abundances derived, for
example, via Lick indices \citep[see][]{BRO90}.

As a preliminary approach, and in order to assess how the inferred abundances 
depend on the adopted calibration, we use two different empirical calibrations.
On the one hand, we adopt the ``inflected" $(g-z)_\mathrm{ACS}$ vs. $[Fe/H]$ relation presented by 
B10 and, on the other, the ``broken line" $(g-i)$ vs $[Z/H]$ 
relation given by U12. The first one relies mostly on Galactic GCs 
and includes some high $[Z/H]$ globulars from NGC\,4486 and NGC\,4472. In turn,
the second  relation stands on a compilation of the literature and spectroscopic
observations of the Calcium triplet lines.

We remark that the upper abundance value given in Table\,5, corresponding
to the B10 calibration, is just a formal extrapolation and, in what
follows (e.g. model fits), we adopt an upper cut-off at $[Z/H]=0.40$. The 
same comment holds for the U12 calibration that only reaches solar abundances 
and is extrapolated up to $[Z/H]=0.33$. With this caveat in mind, we note that 
only a small fraction of the GC candidates in our sample seem to have 
abundances higher than solar.

\section{Globular cluster candidates}
\label{GCs}

\begin{figure}
\includegraphics[width=\hsize]{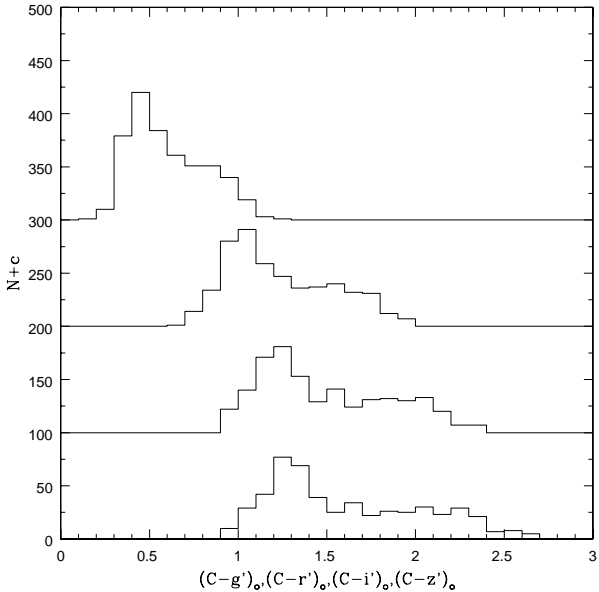}
\caption{ Colour histograms (from top to bottom): $(C-g')o$, $(C-r')o$, $(C-i')o$, $(C-z')o$
 for all the objects listed in Table 3. The histograms have been shifted arbitrarily in ordinates.}
\label{HISTO1}
\end{figure}

\begin{figure}
\includegraphics[width=\hsize]{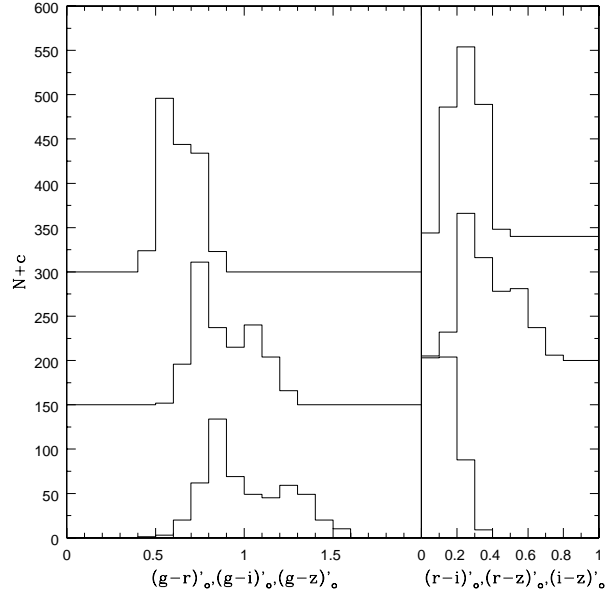}
\caption{ Colour histograms (from top to bottom, left panel): $(g-r)'o$, $(g-i)'o$, $(g-z)'o$ and
(right panel, from top to bottom) $(r-i)'o$, $(r-z)'o$ and $(i-z)'o$ for all the objects listed
 in Table 3. The histograms have beeen shifted arbitrarily in ordinates.}
\label{HISTO2}
\end{figure}

\begin{figure}
\includegraphics[width=\hsize]{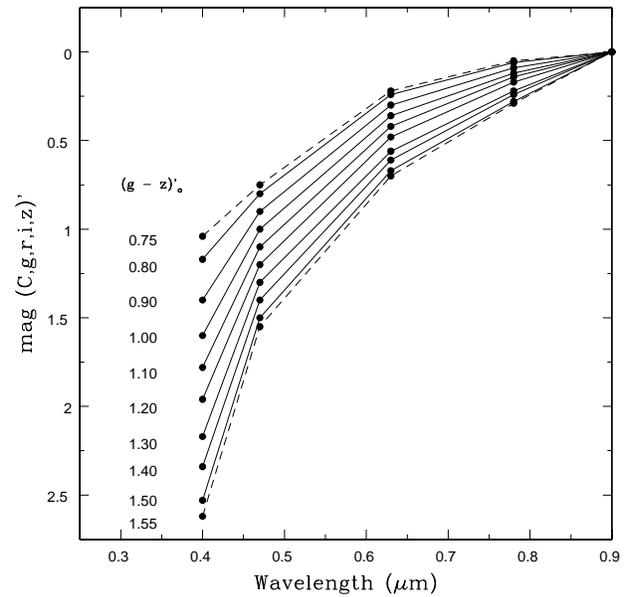}
\caption{Magnitude curves defined by the modal colours listed in
  Table\,\ref{Table_5} from $(g-z)'_0=0.80$ to 1.50 (in 0.1 mag steps;
  continuous lines).  The dashed upper and lower lines correspond to the
  extreme colours $(g-z)'_0=0.75$ and 1.55, respectively. }
\label{CURVESMAG}
\end{figure}

In Figure \ref{HISTO1} and \ref{HISTO2} we display the ten colour histograms
defined in terms of the $Cgriz'$ photometry presented in this paper. These
colour distributions correspond to all the objects listed in Table\,\ref{Table_3}. 
This sample includes a
fraction of field interlopers. In order to decrease the effect of these
objects in our following analysis, we use the template values listed in
Table~\ref{Table_5} in an attempt to identify the genuine GCs. We stress
that the rejected objects, however, have a negligible effect on the
definition of the modal colour-colour relations.  That table defines, at
each $[Z/H]$, a template magnitude curve (or pseudo spectral distribution) as
shown in Figure\,\ref{CURVESMAG}. In this diagram we arbitrarily adopt a
reference magnitude $z'_0=0.0$ mag. These curves assume that GCs are coeval
on the basis of the arguments given before.


The magnitude curves were used to fit the $({\it Cgriz'})_0$ magnitudes of all
the objects listed in Table\,\ref{Table_3}, i.e., we vary the $[Z/H]$ value
by interpolating in Table\,\ref{Table_5} and adopting 0.01 steps in $[Z/H]$
until a given template curve minimizes the sum of the square residuals
at each filter band. This procedure was performed using each of the chemical
abundance scales.

As the results are almost identical in terms of accepted and rejected GC 
candidates, and for illustrative purposes, in this Section we
only show the diagrams corresponding to the adoption of the U12 calibration.

\begin{figure}
\includegraphics[width=\hsize]{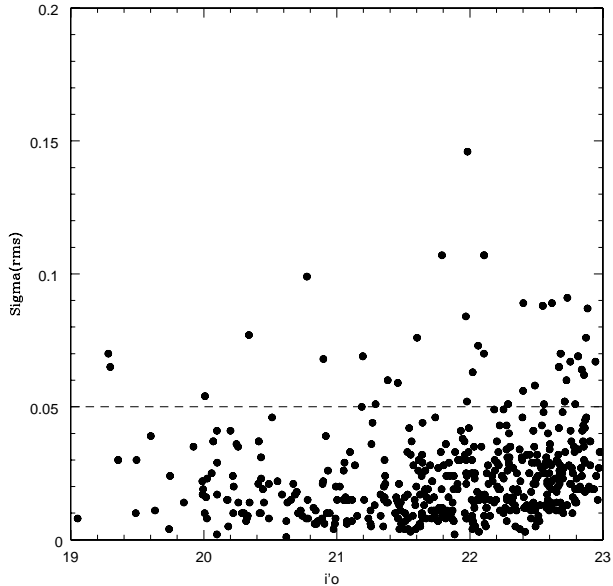}
\caption{Root mean square values corresponding to the magnitude curve fits 
as a function of the $i'_o$ magnitudes for the 521 objects included in 
Table\,3 and adopting the U12 chemical abundance scale. Objects below 0.05 mag 
(dotted line) are considered as GC candidates.}
\label{RESIDUALS_U}
\end{figure}

Figure\,\ref{RESIDUALS_U} shows the overall composite $rms$ (defined by
combining all the magnitude residuals) as a function of the $i'_0$
magnitude.  We adopted this magnitude, in particular, because in this way
all the photometric errors in that diagram are decoupled. The dotted line 
indicates a $rms = 0.05$ mag that we adopt as a ``reasonable'' maximum 
acceptable value to consider a given object as a genuine GC candidate. With 
this criterion, the B10 scale accepts 463 objects as GC candidates and rejects 
58, while the U12 scale leads to 472 and 49, respectively. The difference in the
number of rejected objects arises as a consequence of the upper colour limit
in the B10 calibration (see Table 5). This limit produces $rms$ values larger than
the adopted rejection limit for nine objects with colours redder than $(C-g')_0=0.95$, 
$(r-z)'_0=0.65$, and $(C-z')_0=2.5$.
 

\begin{figure}
\includegraphics[width=\hsize]{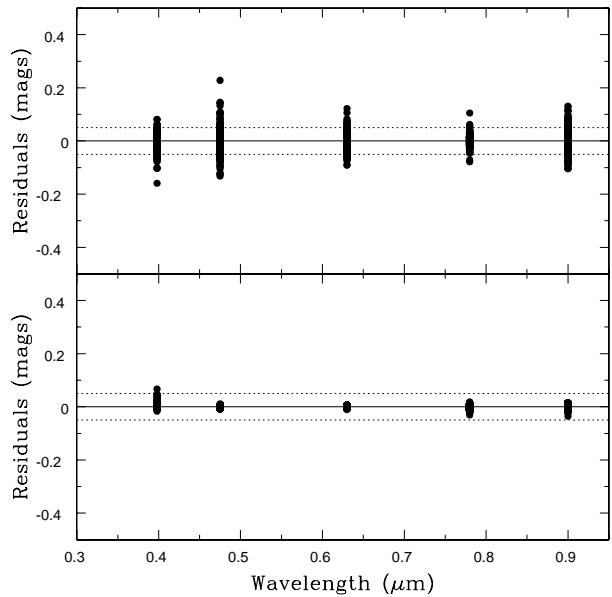}
\caption{\emph{Upper panel:} Magnitude residuals ($C, g',r',i',z'$ bands)
  for the whole sample (521 objects). Dashed horizontal lines indicate
  the adopted $rms = \pm 0.05$ mag limits that define the ``clean'' GC
  sample. \emph{Lower panel:} Expected residuals from Monte Carlo modelling 
  for an age spread of $\pm 2$\,Gyr (see text).}
\label{RESIDUALSWV}
\end{figure}

The upper panel in Figure\,\ref{RESIDUALSWV} shows the
residuals yielded by this procedure at each filter (characterized by their
effective wavelength). These residuals do not show any systematic trend.


As a test of the effects due to an eventual range in the GC ages, we
generated a Monte Carlo model that includes a {\bf spread} of $\pm 2$\,Gyr on
top of the colour-abundance relations given in Table\,\ref{Table_5}. The
differential magnitude variations as a function of age were obtained using
the \citet{MAR05} models and adopting the 12\,Gyr models as the reference
age.

The output colours, including age variations but not photometric errors,
were then fit with the GC templates, as in the case of the observed
clusters, yielding the residuals displayed in the lower panel of
Figure\,\ref{RESIDUALSWV}. The behaviour of these residuals indicates
that age effects have no important impact on the fits as they are
considerably smaller than the photometric errors. On the other hand, the
same Monte Carlo models show that, with our photometric errors, the input
$[Z/H]$ values can be recovered through the magnitude curve fit with an
$rms\approx 0.15$ dex.

The number of rejected objects is consistent with previous
results based on GMOS spectroscopy. In this last case, the photometric
criteria adopted to define a GC candidate typically yield a contamination
level of 10 percent once these candidates are observed spectroscopically
\citep[see][]{FAI11}.

\begin{figure}
\includegraphics[width=\hsize]{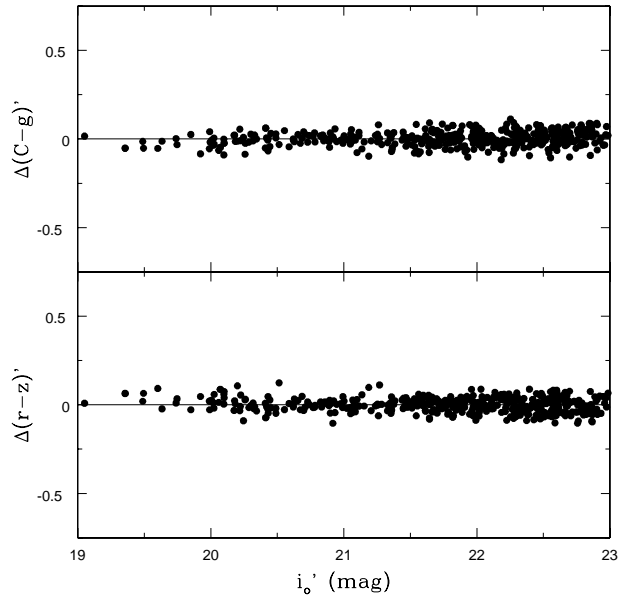}
\caption{Fit residuals in the $(C-g)$ and  $(r-z)'_0$ colours as a function 
of $i'_0$ magnitude for all the GC candidates adopting the U12 calibration.}
\label{RES2}
\end{figure}

The residuals of the $(C-g')_0$ and $(r-z)'_0$ colours  as a function of the
$i'_0$ magnitude are shown in Figure\,\ref{RES2}. GC candidates
fainter than $i'_0\simeq 20.5$ mag do not show obvious systematic
behaviours. However, small anti-correlated trends are observed for the
brightest objects.

We have been unable to reproduce the residual trends seen for the brightest 
objects just by changing the cluster ages in our models. Neither the 
photometric errors, nor those connected with the inferred shape of the 
multicolour relations, seem adequate explanations for this effect.

\begin{figure}
\includegraphics[width=\hsize]{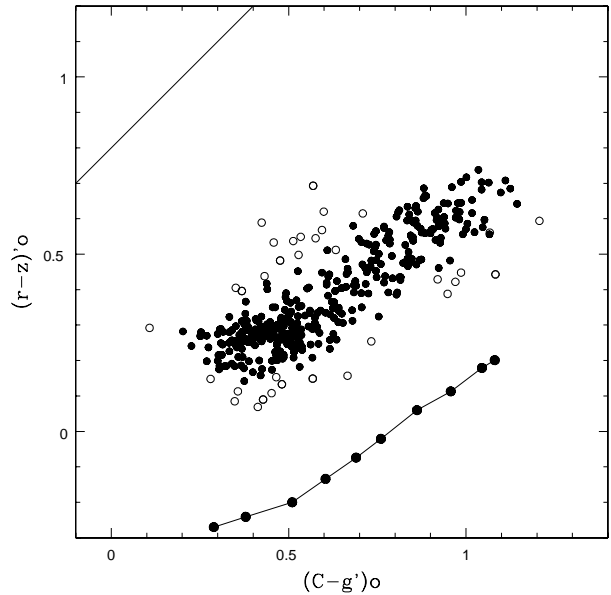}
\caption{$(r-z)'_0$ vs. $(C-g')_0$ colours showing accepted \emph{(filled
dots)} and rejected \emph{(open dots)} CG candidates adopting the U12 
calibration and after applying the $rms > 0.05$ mag cleaning criterion. The 
straight line at upper left has a unity slope. The lower curve corresponds to 
colours given in Table\,\ref{Table_5} shifted by $-0.5$ mag in $(r-z)'_0$.}
\label{RZCG}
\end{figure}

\begin{figure}
\includegraphics[width=\hsize]{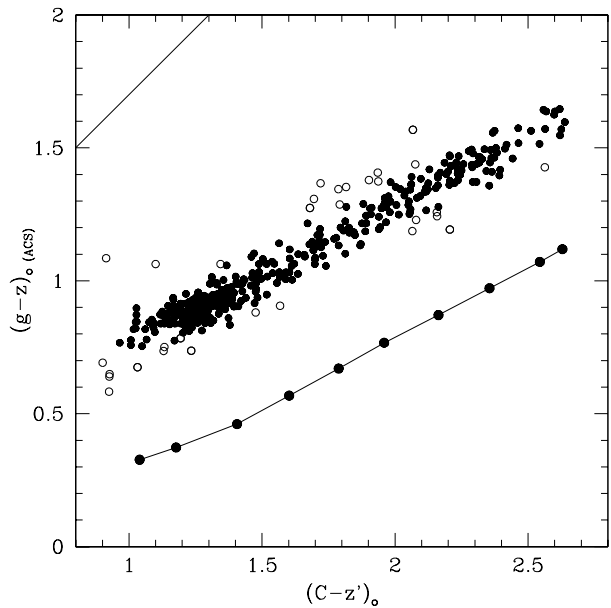}
\caption{${(g-z)_0}_{\mathrm{(ACS)}}$ vs $(C-z')_0$ colours for the accepted
\emph{(filled dots)} and rejected \emph{(open dots)} GC candidates adopting 
the U12 calibration. The straight line at upper left has a unity slope. The 
lower curve corresponds to colours given in Table\,\ref{Table_5} shifted by 
$-0.5$ mag in ${(g-z)_0}_{\mathrm{(ACS)}}$.}
\label{GZACSCZ}
\end{figure}

The clean GC sample, and the rejected objects, are shown in 
Figures\,\ref{RZCG} and \ref{GZACSCZ}, where we transformed $(g-z)'_0$ to
${(g-z)_0}_{\mathrm{(ACS)}}$ through Equation\,\ref{eqn2}.  Both diagrams
show non linear behaviour with detectable changes in their slopes at
${(C-g')_0} \approx 0.5$ and ${(C-z')_0} 
\approx 1.4$, respectively. In particular, Figure\,\ref{GZACSCZ} can be
compared with figure 16 in \citet{CHI11}, which displays
${(g-z)_0}_{\mathrm{(ACS)}}$ as a function of the optical-infrared $(g-K)$
index for GCs in NGC\,4486 and NGC\,4649. Our own reading of that diagram
 indicates that GCs in these galaxies show  changes in the
 colour-colour slope at $(g-K) \approx2.90$ and $(g-K) \approx 3.5$. In 
 turn, GCs bluer than the first of these colours span a $(g-z)_\mathrm{ACS}$ range that covers
 from $ \approx $ 0.8 to  $ \approx $ 1.0 , i.e., the region where we detect a change in the
 colour-colour slope in Figure\,\ref{GZACSCZ}.


\section{Inferring the GC $[Z/H]$s through the template pseudo-continuums}
\label{INVERSE}
 In this section we make an attempt to recover the GCs $[Z/H]$ distribution
 using their integrated colours. A similar approach is presented, for example,
 by \citet{BLA12} (see their figure 11, right panel) for the case of the 
 NGC\,1399 clusters. Instead of using a single integrated colour, in this work 
 we obtain the $[Z/H]$ values for the GC candidates from the ``template curve" 
 fitting described in the previous section. We infer two chemical abundance 
 distributions by alternatively adopting the B10 or U12 calibrations.

 As discussed in several papers, the GC colour bimodality in NGC\,4486 is
 better defined for GCs fainter than $T_{1_0}\approx 21.0$ mag \citep[see, 
 for example, \citetalias{FFG07} or][]{HAR09} and, accordingly, we split our
 sample in two groups: GCs candidates with $T_1$ from 19.0 to 21.0 mag and 
 $T_1$ from 21.0 to 23.0 mag (i.e. $\sim$0.2 mag brighter than the turn-over 
 of the integrated GC luminosity function).

\begin{figure}
\includegraphics[width=\hsize]{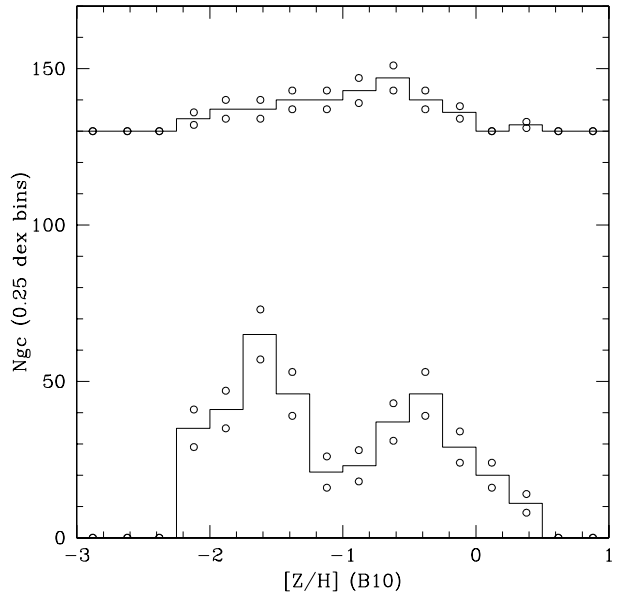}
\caption{Inferred chemical abundance distributions adopting the B10 calibration
 for GC candidates brighter than $T_1=21.0$ mag (shifted upwards by 130 units) 
 and for GC candidates with $T_1$ from 21.0 to 23 mag. Open dots correspond to 
 the formal counting uncertainties for each bin.}
\label{HISTOZHB}
\end{figure}

\begin{figure}
\includegraphics[width=\hsize]{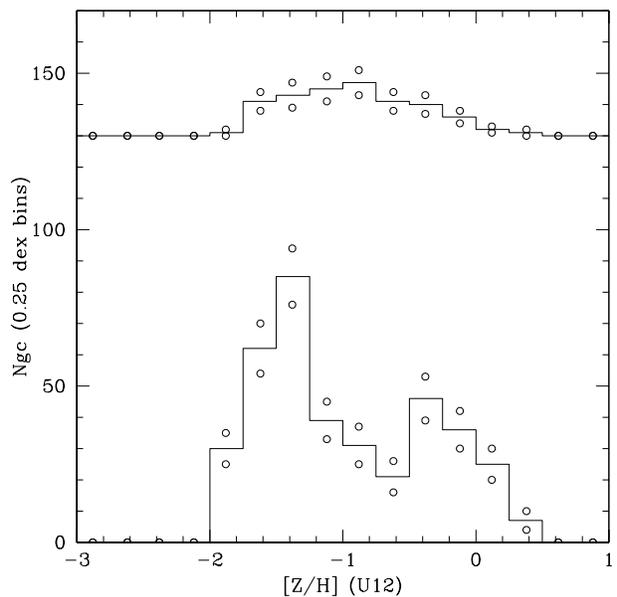}
\caption{Inferred chemical abundance distributions adopting the U12 calibration
 for GC candidates brighter than $T_1=21.0$ mag (shifted upwards by 130 units) 
 and for GC candidates with $T_1$ from 21.0 to 23 mag. Open dots correspond to 
 the formal counting uncertainties for each bin.}
\label{HISTOZHU}
\end{figure}

 Figure\,\ref{HISTOZHB} and Figure\,\ref{HISTOZHU} exhibit the resulting 
 $[Z/H]$ distributions adopting the B10 and U12 calibrations respectively and 
 for the two GC groups. In both diagrams the brightest clusters show distinct 
 behaviour  compared with the fainter counterparts. For the brighter GCs the 
 abundance distributions seem broad and unimodal.

 In turn, independent of the calibration, GCs fainter than $T_1=21$ mag 
 show a clear bimodality although the histograms exhibit differences
 in their shapes and in the position of the low and high abundance peaks (that
 do depend on the adopted calibration).

 A simple Gaussian analysis (using the RMIX routine) indicates that, for 
 these clusters, and for both calibrations, two Gaussian fits are strongly 
 preferred over a single Gaussian fit, leading to $\chi^{2}$ values about 
 3 to 5 times smaller. For the case of the B10 calibration ($\chi^{2}=1.87$)
 we obtain mean $[Z/H]$ values of $-1.66$ and $-0.37$ with dispersions of 
 0.32 and 0.39 dex for the blue (55 percent of the population) and red GCs, 
 respectively. Alternatively, adopting the U12 calibration ($\chi^{2}=0.95$), 
 these parameters become $[Z/H]$ of $-1.42$ and $-0.28$, with dispersions of 
 0.29 and 0.30 dex for the blue GCs (61 percent of the population) and red GC 
 populations, respectively.

As these Gaussians have comparable dispersions, we also attempted a 
homoscedastic KMM test that, in both cases, indicates that the probability for 
a single Gaussian fit to represent the $[Z/H]$ distributions is practically 
null.

The situation of GCs brighter than $T_1=21.0$ mag is not so clear. Both the 
B10 and U12 calibrations lead to broad unimodal distributions that show a different 
degree of skewness.
    
The results shown in Figure\,\ref{HISTOZHB} and Figure\,\ref{HISTOZHU} are 
different, but not necessarily in conflict, with those presented by 
\citet{BLA12} for the central region of NGC\,1399 where red GCs are clearly 
the dominant subpopulation. These authors find a broad unimodal distribution 
with a single peak ($-0.3$ dex) and a tail extending towards lower chemical 
abundances (see their figure\,11, right panel). In turn, our GCs field is 
located at a larger galactocentric distance, where the relative number of blue 
globulars is considerably larger making more evident the presence of a low 
chemical abundance peak.

\section{Modelling the Globular cluster colour histograms}
\label{MODELL}

Two key issues in modelling the GC colour histograms are the adopted
integrated colours vs. chemical abundance calibration and the assumed
chemical abundance distribution of the clusters.  In general, most papers
have relied on a single colour-abundance calibration and also on a single GC
colour histogram fit.  In this work, we attempt a simultaneous fit to the
ten colour histograms that can be defined in terms of the ${\it Cgriz'}$
magnitudes, adopting the (inverse) quality fit indicator $\chi^{2}$, given in
\citet*{COT98}.  Finally, we identify the best fit parameters (that define
the GC chemical abundance distribution) as those that yield a minimum value
for the sum of the individual $\chi^{2}$ indices of the ten colour
histograms (adopting the same colour bin for all the histograms: 0.10 mag).

Monte Carlo model histograms were obtained following the same procedure
explained in \citet{FVF12}. First, we generate ``seed'' globulars with 
chemical abundances $Z$ (within the range $Z_\mathrm{i}$ to $Z_\mathrm{max}$)
whose numbers are controlled by a given distribution function 
$N_\mathrm{GC}(Z)$.

After trying different simple distributions, we concluded, as in FFG07, that
a double exponential dependence:

\begin{equation}
\label{eqn4}
  N_\mathrm{GC}(Z) \approx \exp (-Z/Z_\mathrm{s})
\end{equation}
\noindent 
(where $Z_{s}$ is the scale length corresponding to the blue or red GC
subpopulation) is the simplest function that allows a fit to the colour
histograms based on a {\it minimum} number of free parameters. Formally, this approach
requires seven parameters: the ratio of blue to red clusters, and for each
GC subpopulation the $Z_\mathrm{i}$ and $Z_\mathrm{max}$ values as well as
the chemical scale lenght $Z_{s}$. In fact, the free parameters were reduced to
five, as the minimum chemical abundance of the blue GCs subpopulation,
as well as the maximum chemical abundance of the red GCs, were set as the
lowest and upper values in the B10 and U12 calibrations.

The integrated colours for each GC were obtained by linear
interpolation, using the logarithmic abundance as argument, in
Table\,\ref{Table_5}.
 
For each synthetic cluster we also generate an apparent magnitude $g'_0$,
adopting a Gaussian integrated luminosity function and according to the
parameters given by \citet{VILL10}. These magnitudes were used as input in
Table\,\ref{Table_2} in order to model (also Gaussian) observing errors that
were added to each colour.  Given the relatively short range in apparent
magnitude, we do not include an explicit dependence of chemical abundance
with brightness for the blue globulars \citep[i.e., the ``blue tilt''
effect; see, for example,][]{HAR09}.

The parameters that define the chemical abundance distributions and provide
the best overall fit to the ten colour histograms in each case, are listed
in Table\,\ref{Table_6}, and the corresponding individual and cumulative 
$\chi^{2}$ indices are given in Table\,\ref{Table_7} and Table\,\ref{Table_8}. 

Even though both models lead to $N(GC)$ values within $\approx$ 1.5 times the 
formal counting errors of each histogram bin, the U12 calibration yields 
better fits in terms of the cumulative quality index. We remark that this statement is
valid only if the bi-exponential parametrization of the chemical abundance
is accepted.

Each of the parameters listed in Table\,\ref{Table_6} has an associated 
uncertainty which we define as the parameter variation that leads to a 
decrease of the fit quality, indicated by an increase of ten percent above the 
mimimum total $\chi^{2}$ value in the five free parameters space.

Following this, we get typical uncertainties of $\pm$ 10 GCs for each 
subpopulation; for the blue GCs: $\pm$ 0.01 $Z_{\odot}$ in $Z_{sB}$ and 
$\pm$ 0.05 $Z_{\odot}$ in $Z_\mathrm{max}$; and  for the red GCs: $\pm$ 0.06 
$Z_{\odot}$ in $Z_{sR}$ and $\pm$ 0.02 $Z_{\odot}$ in $Z_{i}$.

\begin{figure}
\includegraphics[width=\hsize]{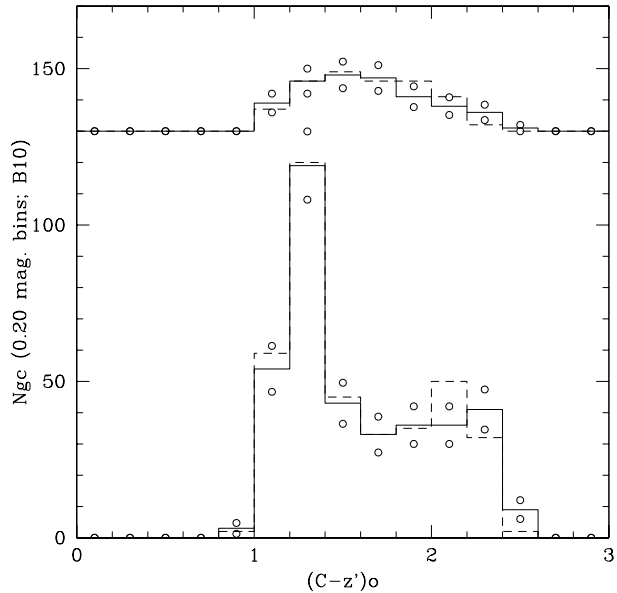}
\caption{GC $(C-z')_0$ colour distributions adopting the B10 calibration 
(solid lines) for GC candidates brighter than $T_1=21.0$ mag (shifted upwards 
by 130 units) and for GC candidates with $T_1$ from 21.0 to 23 mag. Open dots 
correspond to the formal counting uncertainties for each bin. Dashed lines 
correspond to the best fit models given in Table\,6.}
\label{HISTOCZMODB}
\end{figure}

\begin{figure}
\includegraphics[width=\hsize]{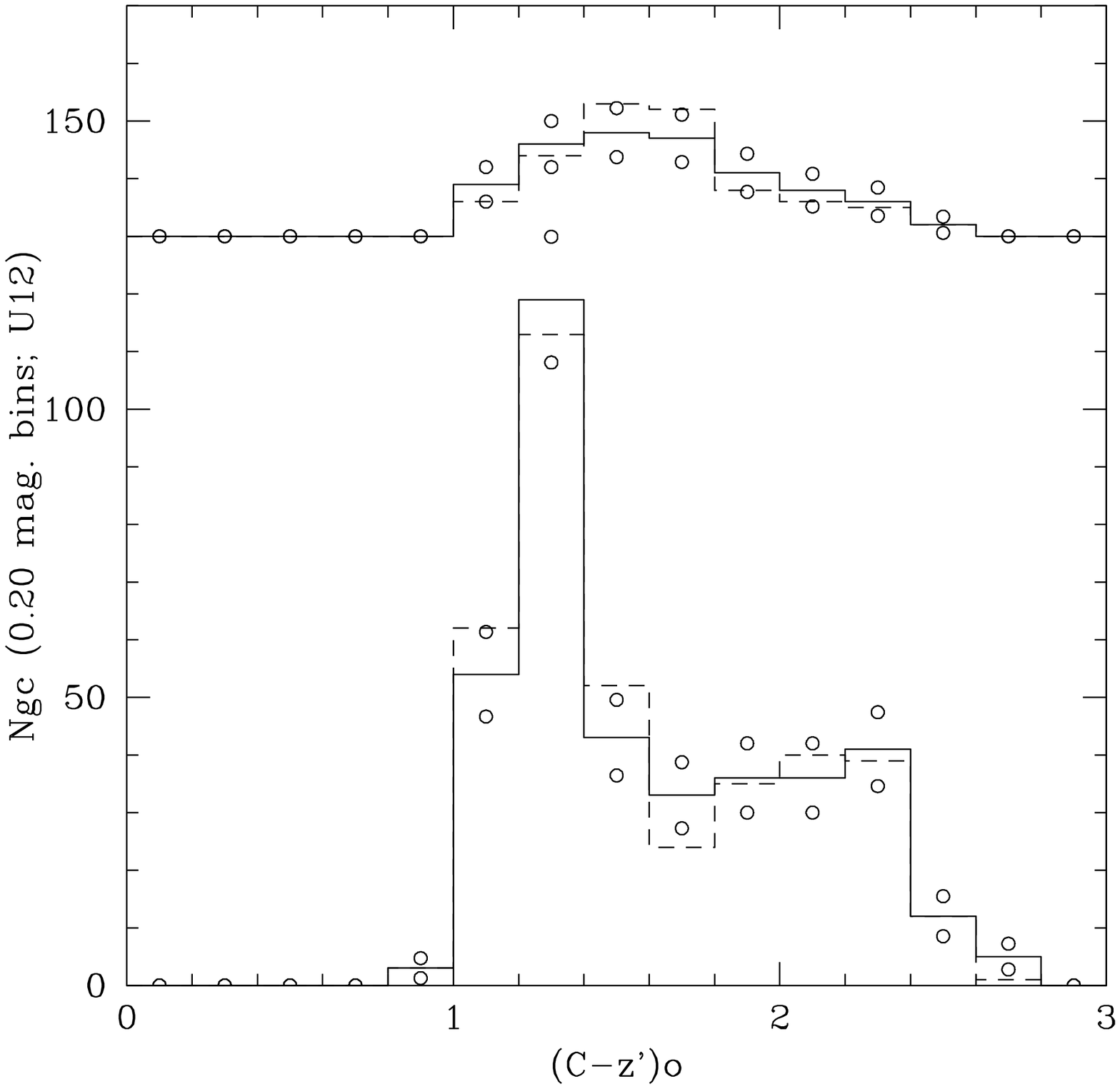}
\caption{GC $(C-z')_0$ colour distributions adopting the U12 calibration 
(solid lines) for GC candidates brighter than $T_1=21.0$ mag (shifted upwards 
by 130 units) and for GC candidates with $T_1$ from 21.0 to 23 mag. Open dots 
correspond to the formal counting uncertainties for each bin. Dashed lines 
correspond to the best fit models given in Table\,6.}
\label{HISTOCZMODU}
\end{figure}

The histograms corresponding to $(C-z')_0$ colour are depicted in 
Figure\,\ref{HISTOCZMODB} and Figure\,\ref{HISTOCZMODU}. We note that this 
index is $\approx 1.6$ and $\approx 2.0$ times more sensitive to metallicity 
than $(C-T_1)$ and $(g-z')$, respectively. This allows the adoption of a larger
colour bin (0.2 mag) thereby reducing the sampling noise. Both histograms exhibit 
a colour ``valley" at  $(C-z')_0\approx1.7$, i.e., 0.3 mag. redder than the colour where a
change in the colour-colour relation is detectable (see Section\,5). The 
models show that both GC populations overlap in the color range 
$(C-z')_0= 1.30 - 1.70$. In this range, some 20 to 25 percent of the total 
number of clusters belong to the ``blue tail" of the red GC population.
 We note that the $(C-z')_0$ colour of the valley corresponds to ${(g-z)_0}_{\mathrm{(ACS)}} \approx1.1$ to
$1.2$, i.e., coincident with the colour region where \citet{CHI11} claim the presence
of a ``wavy feature" in the $(g-z)_\mathrm{ACS}$ vs. $(g-K)$ relation.

\begin{table}
\center
\caption{Parameters of the chemical abundance distributions giving the best overall fit to the colour histograms
for Globular Clusters fainter than  $T_1=21$.}
\label{Table_6}
\begin{tabular}{|cccc|}                                 
\hline
\multicolumn{4}{c}{Adopting the B10 calibration:}\\
$N_b = 200$ & $Z_{\mathrm{s}_b} = 0.02~Z_{\odot}$ & $Z_{\mathrm{i}} = 0.007~Z_{\odot}$ &
$Z_{\mathrm{max}} = 1.0~Z_{\odot}$ \\
$ N_r = 178$ & $Z_{\mathrm{s}_r} = 0.50~Z_{\odot}$ & $Z_{\mathrm{i}} = 0.050~Z_{\odot}$ & $
Z_{\mathrm{max}} = 2.5~Z_{\odot}$\\
\noalign{\smallskip}
\multicolumn{4}{c}{Adopting the U12 calibration:}\\
$N_b = 220$ & $Z_{\mathrm{s}_b} = 0.03~Z_{\odot}$ & $Z_{\mathrm{i}} = 0.014~Z_{\odot}$ &
$Z_{\mathrm{max}} = 1.0~Z_{\odot}$ \\
$ N_r = 160$ & $Z_{\mathrm{s}_r} = 0.65~Z_{\odot}$ & $Z_{\mathrm{i}} = 0.050~Z_{\odot}$ & $
Z_{\mathrm{max}} = 2.5~Z_{\odot}$\\

\hline
\end{tabular}
\end{table}


\begin{table*}
\center
 \begin{minipage}{185mm}
  \caption{GC colour histograms fit $\chi^{2}$ adopting the B10 
calibration. First line, GCs brighter than $T_1=21.0$ mag; second line: GCs 
fainter than $T_1=21.0$ mag.} 
\label{Table_7}
\begin{tabular}{@{}ccccccccccc@{}}                                 
\hline
$(C-g)_0'$ & $(C-r){_0}'$ & $(C-i){_0}'$ & $(C-z){_0}'$ & $(g-r)_0'$ & $(g-i){_0}'$ & $(g-z)_0'$ & $(r-i)_0'$ &$(r-z)_0'$ &$(i-z)_0'$ & Cumulative $\chi^{2}$\\
\hline
  0.85  & 0.64 & 0.90 & 0.74 & 0.61 & 0.92 & 0.41 & 0.08 & 1.41 & 1.42 &   7.98 \\
  0.56  & 1.18 & 1.07 & 1.13 & 0.84 & 1.29 & 1.62 & 0.41 & 1.76 & 1.60 &  11.46 \\
\hline
\end{tabular}
\end{minipage}
\end{table*}
\begin{table*}
\centering
 \begin{minipage}{185mm}
  \caption{GC colour histograms fit $\chi^{2}$ adopting the U12 
calibration. First line, GCs brighter than $T_1=21.0$ mag; second line: GCs 
fainter than $T_1=21.0$ mag.} 
\label{Table_8}
\begin{tabular}{@{}ccccccccccc@{}}                                 
\hline
$(C-g)_0'$ & $(C-r){_0}'$ & $(C-i){_0}'$ & $(C-z){_0}'$ & $(g-r)_0'$ & $(g-i){_0}'$ & $(g-z)_0'$ & $(r-i)_0'$ &$(r-z)_0'$ &$(i-z)_0'$ & Cumulative $\chi^{2}$\\
\hline
 0.62  & 0.56 & 0.61 & 0.59 & 0.62 & 0.59 & 0.49 & 0.11 & 0.58 & 0.37 &   5.27 \\
 0.62  & 0.90 & 0.61 & 0.80 & 1.15 & 0.56 & 0.95 & 0.44 & 0.77 & 0.62 &  7.42 \\
\hline
\end{tabular}
\end{minipage}
\end{table*}

 The GCs $[Z/H]$ distributions inferred from the photometric observations can 
 be compared with those arising in the models that deliver the best fits to 
 the ten observed colour histograms and whose parameters are listed in 
 Table~\ref{Table_6}.
 
\begin{figure}
\includegraphics[width=\hsize]{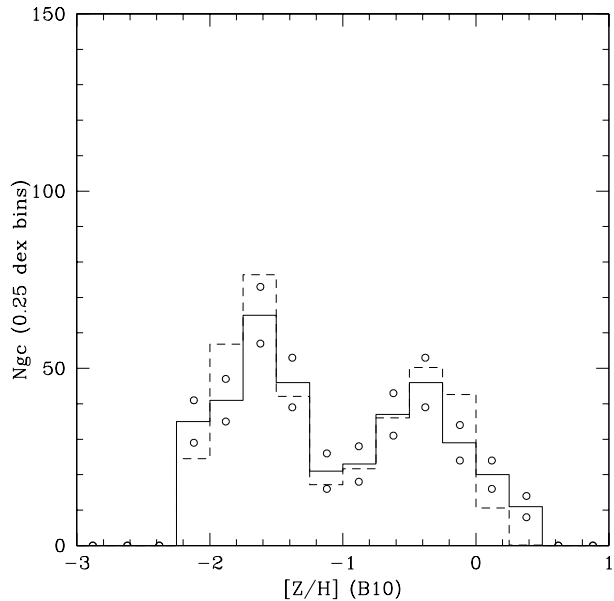}
\caption{Comparison between the inferred $[Z/H]$ GC distributions (solid line 
 histogram) for GCs fainter than $T_1= 21$ mag, and that 
 derived from the colour histogram fits adopting the B10 calibration (dashed 
 lines). Open dots are the formal counting uncertainties in each bin.}
\label{HISTOZHMODB}
\end{figure}

\begin{figure}
\includegraphics[width=\hsize]{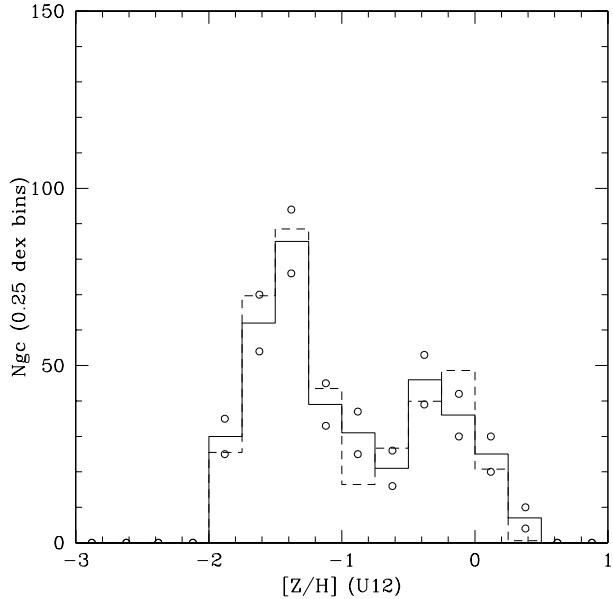}
\caption{Comparison between the inferred $[Z/H]$ GC distributions (solid line 
 histogram) for GCs fainter than $T_1=21$ mag, and that derived 
 from the colour histogram fits adopting the U12 calibration (dashed lines). 
 Open dots are the formal counting uncertainties in each bin.}
\label{HISTOZHMODU}
\end{figure}

 Figures\,\ref{HISTOZHMODB} and~\ref{HISTOZHMODU} display such a comparison. 
 For clusters fainter than $T_1=21.0$ mag the models based on the B10 
 calibration yield mean $[Z/H]$ of $-1.69$ and $-0.47$ while the adoption of 
 the U12 calibration gives somewhat larger values, $-1.46$ and $-0.36$. These 
 mean abundances are close to, but different from, the values arising from the
 simple Gaussian analysis presented before. 

 {\bf The fits to the brighter clusters lead to an ambiguous situation given the
 small number of GCs and the absence of definite peaks that prevents us from
 obtaining robust results}.

\section{Conclusions}
\label{CONCLU}

 This paper presents a self consistent multicolour ${\it Cgriz'}$ grid 
 including 100 points, each with a typical uncertainty of $\approx \pm 0.012$ 
 mag, and based on GC candidates in NGC\,4486. This grid, once calibrated in 
 terms of two different colour-metallicity relations, has been used to infer 
 the GCs chemical abundances from ${\it Cgriz'}$ photometric data and to 
 perform a comparison with simple models.

 The main results are:

\renewcommand{\labelenumi}{\arabic{enumi})}   

\begin{enumerate}

\item The multicolour ${\it Cgriz'}$ relations show different degrees of non 
linearities. This is more evident in those colours involving the $C$ filter. 
Non linearities of this kind had been previously reported by \citet{BLA12} in 
their study of the central regions of NGC\,1399.
    
\item The inferred $[Z/H]$ distributions for GCs fainter than $T_1=21.0$ mag 
are bimodal either adopting the ``inflected" B10 or the ``broken line" U12 
colour-abundance calibrations.

\item The model fits based on a double exponential dependence of the number of 
clusters with chemical abundance $Z$ both for the blue and red GCs provide a 
good representation of the GC integrated colour histograms and of their 
inferred chemical abundance distributions.
    
\item For the brightest clusters the abundance distributions 
appear broad and skewed and we do not reach a definite conclusion regarding
the presence of bi-modality. These objects leave systematic colour residuals
from the template magnitude curves that cannot be easily accounted for. In a
speculative way, these residuals might indicate the presence of multi-stellar
populations similar to those found in systems with comparable absolute magnitudes 
(e.g. \citealt{BOE12}).

\item Adequate two-colour diagrams, such as $(C-g')_0$ vs. $(r-z)'_0$ or 
$(C-z')_0$ vs. $(g-z)'_0$, show changes in the colour-colour slopes. These
changes are detectable, for example, at $(C-z')_0=1.4$. On the other hand, the
bi-exponential modelling, adopting the B10 or U12 colour-chemical abundance relations,
show that 88 and 82 percent, respectively, of the ``blue" GCs are bluer
than that colour. This indicates that the ``blue" and ``red" GC sub-populations exhibit
distinct colour-colour relations with a transition zone possibly between $(C-z')_0=1.40$
and $(C-z')_0=1.60$ (where both sub-populations overlap to some degree).

 The results presented in this paper suggest that the origin of the GC colour
 bimodality has its roots in a real bimodality of their chemical abundance distributions, 
 and are consistent with the spectroscopic analysis of GCs in roughly half
 of the sample of eight galaxies studied by U12 and also by \citet{BRO12} in NGC\,3115.

\end{enumerate}
        
\section*{Acknowledgements} 
JCF acknowledges Prof. Luc\'ia Send\'on (Director) and the staff of the
Planetario ``Galileo Galilei'' (Buenos Aires) for their hospitality. We also
thank the Referee, Dr. John Blakeslee, for careful reading and commnents that improved the
original version.

 AVSC acknowledges finantial support from Agencia de Promoci\'on Cient\'ifica y
Tecnol\'ogica of Argentina (BID AR PICT 2010-0410).  This work was supported
by grants from La Plata National University, Agencia Nacional de Promoci\'on
Cient\'ifica y Tecnol\'ogica, and CONICET (PIP-200801-1611 and PIP-2009-0712),
Argentina. D.G. gratefully acknowledges support from the Chilean 
BASAL   Centro de Excelencia en Astrof\'isica
y Tecnolog\'ias Afines (CATA) grant PFB-06/2007.

\label{lastpage}

\end{document}